\documentclass[aps,pra,twocolumn,reprint]{revtex4}
\usepackage{amsfonts, amssymb, amsmath,graphicx}
\usepackage{subfigure}
\usepackage{multirow}
\usepackage{dcolumn}
\usepackage{bm}
\usepackage{color}

\begin{document}

\title{Momentum-space Harper-Hofstadter model}

\author{Tomoki Ozawa}
\author{Hannah M. Price}
\author{Iacopo Carusotto}
\affiliation{INO-CNR BEC Center and Dipartimento di Fisica, Universit\`a di Trento, I-38123 Povo, Italy}%

\date{\today}

\def\simge{\mathrel{%
         \rlap{\raise 0.511ex \hbox{$>$}}{\lower 0.511ex \hbox{$\sim$}}}}
\def\simle{\mathrel{
         \rlap{\raise 0.511ex \hbox{$<$}}{\lower 0.511ex \hbox{$\sim$}}}}
\newcommand{\feynslash}[1]{{#1\kern-.5em /}}
\newcommand{\iac}[1]{{\color{red} #1}}
\newcommand{\tom}[1]{{\color{blue} #1}}

\begin{abstract}
We show how the weakly trapped Harper-Hofstadter model can be mapped onto a Harper-Hofstadter model in momentum space. In this momentum-space model, the band dispersion plays the role of the periodic potential, the Berry curvature plays the role of an effective magnetic field, the real-space harmonic trap provides the momentum-space kinetic energy responsible for the hopping, and the trap position sets the boundary conditions around the magnetic Brillouin zone. Spatially local interactions translate into nonlocal interactions in momentum space: within a mean-field approximation, we show that increasing interparticle interactions leads to a structural change of the ground state, from a single rotationally symmetric ground state to degenerate ground states that spontaneously break rotational symmetry.
\end{abstract}

\maketitle

\section{Introduction}

Since the discovery of the key role played by the nontrivial topology of energy bands in quantum Hall systems~\cite{Thouless1982, Kohmoto1985}, topological states of matter have attracted considerable attention across many areas of physics~\cite{Hasan2010,Qi2011}.
While originally observed in solid-state systems~\cite{Klitzing1980, Konig2007, Hsieh2008}, a variety of geometrically and topologically nontrivial states have now also been realized in ultracold atomic gases~\cite{Tarruell2012, Struck2013, Atala2013, Aidelsburger2013, Miyake2013, Aidelsburger2014, Jotzu2014, Kennedy2015} and photonics~\cite{Wang2009, Rechtsman2013a, Rechtsman2013b, Hafezi2013}. 
Most of the topological models considered so far are based on real-space lattices, while much less is known about the nontrivial topological features of {\em momentum-space} lattices~\cite{Cooper2012,Scaffidi2014}.

In this paper, we propose a simple way to realize a momentum-space Harper-Hofstadter (HH) model, namely, a two-dimensional tight-binding lattice with an effective magnetic field in {\it momentum space}. The lattice potential is provided by the periodic energy-band dispersion of the underlying real-space system, while a harmonic trap in real space is crucial in providing the hopping in momentum space~\cite{Price2014}. Finally, the role of the magnetic field is played by the Berry curvature, a geometrical property of the energy band~\cite{Bliokh2005, Berry1984}.
While the momentum-space HH model has the same local structure as the usual real-space HH model~\cite{Harper1955, Hofstadter1976}, its {\em global} topology is very different: the real-space model can have any extension in space, while the momentum-space model is intrinsically restricted to the magnetic Brillouin zone (MBZ), which has the topology of a two-dimensional torus. 

As a first application of the momentum-space HH model, we extend our previous work~\cite{Price2014} and determine the ground state of a trapped noninteracting real-space HH model beyond the small-flux limit, in regimes where the energy dispersion of the lowest energy band is no longer negligible and toroidal Landau levels are strongly deformed. As the boundary conditions around the momentum-space torus are controlled by the spatial position of the harmonic trap in real space, a momentum-space version of Laughlin's charge pumping {\em Gedankenexperiment}~\cite{Laughlin1981} is anticipated.

We then move to the interacting case: while the interacting HH model in a spatially homogeneous geometry has already been the subject of several works~\cite{Balents2005,Powell2010,Powell2011}, not much is yet known about the unexpectedly rich physics that is introduced by the harmonic confining potential~\cite{Harper2014}. Most remarkably, as a function of the strength of interactions, we find transitions that spontaneously break rotational symmetry and lead to degenerate ground states. 

Since the real-space HH model has been realized in ultracold atomic gases~\cite{Aidelsburger2013, Miyake2013, Aidelsburger2014}, photonic devices~\cite{Hafezi2013}, and solid-state superlattices~\cite{Dean2013, Yu2014}, the addition of a controlled harmonic trapping potential can be a direct extension of existing experiments. Most directly, the observation of a Bose-Einstein condensate in the real-space HH model was recently reported in~\cite{Kennedy2015}: with respect to this, our study highlights the crucial role of the external trap in determining the condensate mode. More generally, our conclusions are expected to hold in any configuration in which the single-particle dispersion has multiple minima as in~\cite{Mueller2006, Lin2011, Parker2013}.

This paper is organized as follows: in Sec. II, we apply the momentum-space formalism~\cite{Price2014} to the weakly trapped noninteracting HH model, and we show how this model maps to the HH model in momentum space. In Sec. III, we give the effective Hamiltonian of the momentum-space HH model and compare the prediction of the momentum-space HH model with the numerical simulation of the original real-space HH model. In Sec. IV, we include on-site interparticle interactions to the model and discuss how the ground state changes its symmetry, showing how this can be understood in terms of the momentum-space HH model. Finally, our conclusions are presented in Sec. V.

\section{Momentum-space formalism}

We first briefly review the momentum-space formalism developed in~\cite{Price2014} before applying this formalism to the harmonically trapped real-space HH model.
We consider the situation where the Hamiltonian is of the form $\mathcal{H} = \mathcal{H}_0 + \frac{1}{2}\kappa r^2$, where $\mathcal{H}_0$ is periodic in space and $\kappa$ is the strength of an additional harmonic potential. In the absence of the harmonic potential ($\kappa = 0$), the Hamiltonian is periodic, and the eigenstates follow Bloch's theorem: $\mathcal{H}_0 e^{i\mathbf{k}\cdot \mathbf{r}}|u_{n,\mathbf{k}}\rangle = E_{n}(\mathbf{k}) e^{i\mathbf{k}\cdot \mathbf{r}}|u_{n,\mathbf{k}}\rangle$, where $n$ is the band index, $\mathbf{k}$ is the quasimomentum, and $|u_{n,\mathbf{k}}\rangle$ is a periodic function with the same periodicity as $\mathcal{H}_0$. The eigenvalue $E_n (\mathbf{k})$ is the band dispersion of the $n$th band. In the presence of a harmonic potential ($\kappa > 0$), the eigenstates no longer satisfy Bloch's theorem, but we can still expand any state in the basis of Bloch states $e^{i\mathbf{k}\cdot\mathbf{r}}|u_{n,\mathbf{k}}\rangle$ because these form a complete basis set. Expanding a state $|\Psi\rangle$ in the Bloch basis as $|\Psi\rangle = \sum_{n,\mathbf{k}}\psi_{n,\mathbf{k}} e^{i\mathbf{k}\cdot\mathbf{r}}|u_{n,\mathbf{k}}\rangle$, the expansion coefficients $\psi_{n,\mathbf{k}}$, which have the physical meaning of the momentum-space wave function, follow the Schr\"odinger-like equation
\begin{align}
	i\frac{\partial}{\partial t} \psi_{n,\mathbf{k}}
	=
	\left[E_{n}(\mathbf{k}) + \frac{\kappa}{2}[i\nabla_\mathbf{k} + \mathcal{A}_{n}(\mathbf{k})]^2\right]
	\psi_{n,\mathbf{k}},
\end{align}
where the Berry connection is $\mathcal{A}_n (\mathbf{k}) \equiv i\langle u_{n,\mathbf{k}} | \nabla_\mathbf{k} | u_{n,\mathbf{k}}\rangle$. In writing this down, we have assumed that the additional harmonic trap is weak enough that the contribution from only a single band $n$ is significant.
Then, the physics of the momentum-space wave function $\psi_{n,\mathbf{k}}$ is described by the effective momentum-space Hamiltonian
\begin{align}
	\tilde{\mathcal{H}}
	=
	E_{n}(\mathbf{k}) + \frac{\kappa}{2}[i\nabla_\mathbf{k} + \mathcal{A}_{n}(\mathbf{k})]^2,
	\label{momentumham}
\end{align}
which is analogous to the Hamiltonian of a charged particle in an external electromagnetic field, where the role of the mass, magnetic vector potential, and the electrostatic potential are replaced by the inverse trapping strength $1/\kappa$, the Berry connection $\mathcal{A}_n (\mathbf{k})$, and the band dispersion $E_n (\mathbf{k})$, respectively. Unless otherwise specified, the simpler term ``momentum'' will be used to indicate the quasimomentum $\mathbf{k}$.

\subsection{Harper-Hofstadter model}

Now we apply this formalism to the weakly trapped real-space HH model and discuss how this model can be mapped onto a momentum-space HH model.
We consider the bosonic HH model on a square lattice with a harmonic potential, described by the following Hamiltonian:
\begin{align}
	&\mathcal{H} =
	-J\sum_{m,n} \left( a_{m+1,n}^\dagger a_{m,n} + e^{i2\pi \alpha m}a_{m,n+1}^\dagger a_{m,n} + \mathrm{H.c.} \right)
	\notag \\
	&\phantom{=}+\frac{1}{2}\kappa \sum_{m,n} \{(m-m_0)^2 + (n-n_0)^2\} a_{m,n}^\dagger a_{m,n},
	\label{hamiltonian}
\end{align}
where $a_{m,n}$ is the bosonic annihilation operator at site $(m,n)$, $J > 0$ is the hopping amplitude, and $\alpha$ is the number of magnetic flux quanta per plaquette. The site-dependent hopping phase corresponds to the magnetic vector potential 
\begin{equation}
(A_x, A_y) = (0, 2\pi \alpha x)
\label{A_magn}
\end{equation}
in the Landau gauge. The harmonic trap has a strength $\kappa$, and the coordinates of its center $(m_0, n_0)$ can take noninteger values.

In the absence of a trap, the HH model with a rational flux $\alpha = p/q$ (with $p$ and $q$ coprime integers) is characterized by $q$ topologically nontrivial energy bands with nonzero Berry curvature and Chern number~\cite{Thouless1982}. The lowest-energy-band dispersion $E(\mathbf{k})$ has minima at $(2\pi \mu/q, 2\pi \nu/q)$ in the extended zone scheme, where $\mu$ and $\nu$ are integers. Of these minima, $q$ are located within the natural $2\pi/q \times 2\pi$ magnetic Brillouin zone of the Landau magnetic gauge.

Then, in the presence of a trap, the real-space model described by (\ref{hamiltonian}) maps, according to~(\ref{momentumham}), in momentum space to a particle of mass $\kappa^{-1}$ moving under a scalar potential given by the HH energy dispersion $E(\mathbf{k})$ and a vector potential given by the Berry connection $\mathcal{A}(\mathbf{k})$. As the HH energy dispersion $E(\mathbf{k})$ is periodic in $\mathbf{k}$, the momentum-space Hamiltonian has the form of a particle moving in a periodic potential in momentum space. 

If the momentum-space mass is sufficiently heavy (i.e., the trap strength $\kappa$ is sufficiently weak), we can consider the {\it tight-binding model} of the {\it momentum-space Hamiltonian}, where the Wannier states localized in each minimum of the dispersion in momentum space are taken as a basis. This leads to a tight-binding model in momentum space in which the Berry curvature acts as an effective magnetic field: the {\em momentum-space HH model}~\footnote{A first discussion of such momentum-space HH models appeared in Scaffidi and Simon~\cite{Scaffidi2014} but was restricted to a special lattice size of $q\times q$ sites. Our analysis presented here holds for general lattice sizes.}.

It is worth stressing that a sizable width of the band is essential to get an efficient localization at the minima of the momentum-space periodic potential $E(\mathbf{k})$: this condition is fulfilled by the real-space HH model for $\alpha=1/q$ with relatively small $q$. This regime is to be contrasted with our previous work~\cite{Price2014}, where we focused on the small $\alpha\ll 1$ case where the energy dispersion of the HH energy bands is effectively negligible.

\subsection{Boundary condition in momentum space}

\begin{figure}[htbp]
\begin{center}
\includegraphics[width=5.5cm]{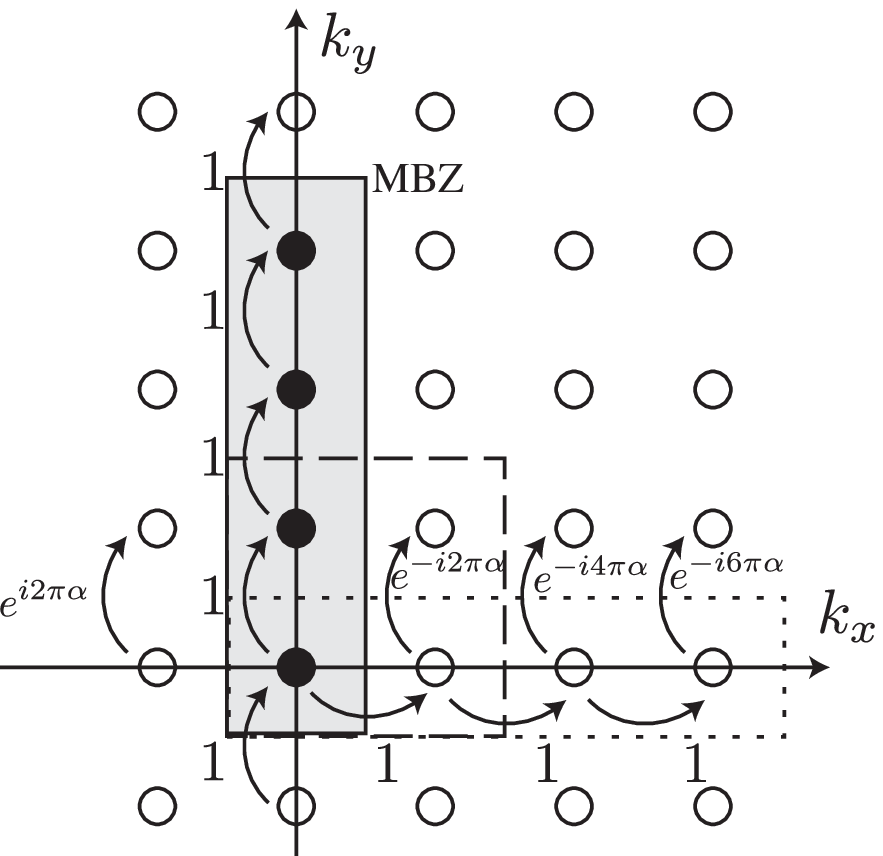}
\caption{Momentum-space HH model for the lowest band of $\alpha = 1/4$ which has $\mathcal{C} = -1$. The MBZ enclosed by a solid line is the natural MBZ corresponding to the magnetic gauge of our choice~(\ref{hamiltonian}). Differently shaped MBZs, described by nonsolid lines, can be obtained by using different magnetic gauges. The hoppings are given up to a constant factor.}
\label{hofstadter}
\end{center}
\end{figure}

The magnetic Brillouin zone in which the momentum-space dynamics takes place is shown in Fig.~\ref{hofstadter}.
It has a rectangular shape of size $2\pi/q \times 2\pi$ and, because of the periodic boundary conditions, a global toroidal topology~\cite{Price2014}. The number of magnetic flux quanta piercing the torus equals the Chern number $\mathcal{C}$ of the band~\cite{Fang2003}. Given the $q$-fold periodicity of the energy dispersion and of the Berry curvature within the magnetic Brillouin zone, the momentum-space plaquette has a square shape of side $2\pi/q$, and the magnetic flux per momentum-space plaquette is $\mathcal{C}/q$~\cite{Scaffidi2014}.

To clearly distinguish the gauge associated with the real-space magnetic vector potential (\ref{A_magn}) from the gauge associated with the Berry connection $\mathcal{A}_n(\mathbf{k})$ in each energy band, we denote the former as the {\em magnetic gauge} and the latter as the {\em Berry gauge}.
Since the Berry curvature has a periodicity of $2\pi/q$ along $k_x$ and $k_y$, we can decompose it into a sum of two terms: an average part, $\bar{\Omega} = q\mathcal{C}/2\pi$, and a periodic part with a vanishing average. For each band (in particular for the lowest one under consideration here), a suitable choice of the Berry gauge can be found such that the Berry connection has the form
\begin{align}
	(\mathcal{A}_x, \mathcal{A}_y) = (0, \bar{\Omega} k_x + \delta \mathcal{A}_y),
	\label{ourberrygauge}
\end{align}
where the contribution of the average Berry curvature $\bar{\Omega}$ to the Berry connection is chosen in the Landau gauge, and the correction term is
\begin{equation}
\delta \mathcal{A}_y \equiv \int_0^{k_x}dk_x^\prime \left[ \Omega(k_x^\prime,k_y) - \bar{\Omega}\right].
\end{equation}
The Berry connection vanishes along the $k_x = 0$ line ($\mathcal{A}_y = 0$). As a result, hopping in the $k_y$ direction along the $k_x=0$ line has no Peierls phase.

It is well known that no periodic and smooth gauge can be found over the surface of a torus pierced by a nonzero total magnetic flux. As a result, the periodicity of the momentum-space wave function $\psi(k_x,k_y)$ is preserved only up to a phase~\cite{AlHashimi2009}:
\begin{align}
	\psi (\mathcal{L}_x, k_y) &= e^{i\phi_x (k_y)} \psi(0, k_y), \notag \\
	\psi (k_x, \mathcal{L}_y) &= e^{i\phi_y (k_x)} \psi(k_x, 0), \label{momentumboundary}
\end{align}
where $\mathcal{L}_x=2\pi/q$ and $\mathcal{L}_y=2\pi$ are the MBZ sizes along $k_x$ and $k_y$, and $\phi_x (k_y)$ and $\phi_y (k_x)$ are smooth transition functions, which characterize the boundary conditions.
Correspondingly, the Berry connections 
$\mathcal{A}_x$ and $\mathcal{A}_y$ at the boundary obey
\begin{align}
	\mathcal{A}_y (\mathcal{L}_x, k_y) &= \mathcal{A}_y (0, k_y) + \partial_{k_y} \phi_x (k_y), \notag \\
	\mathcal{A}_x (k_x, \mathcal{L}_y) &= \mathcal{A}_x (k_x, 0) + \partial_{k_x} \phi_y (k_x), \label{boundarya}
\end{align}
with $\partial_{k_i} \equiv \partial / \partial k_i$.
For our choice~(\ref{ourberrygauge}) of the Berry gauge, the boundary conditions (\ref{boundarya}) impose that the transition functions take the form
\begin{align}
	\phi_x (k_y) &= \theta_x + \int_{0}^{\mathcal{L}_x} dk_x^\prime \int_0^{k_y}dk_y^\prime \Omega(k_x^\prime, k_y^\prime),
	\notag \\
	\phi_y (k_x) &= q\theta_y,
	\label{boundaryexpression}
\end{align}
where $\theta_x$ and $q\theta_y$ are integration constants.

The angles $\theta_{x,y}$ determining the boundary conditions $\phi_x(k_y)$ and $\phi_y(k_x)$ around the momentum-space torus are fixed by the real-space Hamiltonian, as is the case for the energy-band dispersion $E(\mathbf{k})$ and the Berry connection $\mathcal{A}(\mathbf{k})$. For the real-space HH model of Hamiltonian (\ref{hamiltonian}) under consideration here, these can be calculated as follows. 

On a two-dimensional (2D) torus, Berry-gauge-invariant quantities called the (twisted) Polyakov loops $\Phi_{x,y}$ exist~\cite{AlHashimi2009}:
\begin{align}
	\Phi_x (k_y) &\equiv \phi_x (k_y) - \int_{0}^{\mathcal{L}_x} \mathcal{A}_x (k_x^\prime, k_y)dk_x^\prime,
	\label{Poly_y} \\
	\Phi_y (k_x) &\equiv \phi_y (k_x) - \int_{0}^{\mathcal{L}_y} \mathcal{A}_y (k_x, k_y^\prime)dk_y^\prime.
\label{Poly_x}
\end{align}
For our choice of the Berry gauge~(\ref{ourberrygauge}) the Polyakov loops are related to the unknown angles $\theta_{x,y}$ by
\begin{align}
	\Phi_x (k_y) &= \theta_x + \int_{0}^{\mathcal{L}_x} dk_x^\prime \int_0^{k_y}dk_y^\prime \Omega(k_x^\prime, k_y^\prime),
	\notag \\
	\Phi_y (k_x) &= q\theta_y - \int_{0}^{k_x} dk_x^\prime \int_0^{\mathcal{L}_y}dk_y^\prime \Omega(k_x^\prime, k_y^\prime).
	\label{phixphiy}
\end{align}
Thanks to their Berry gauge invariance, the value of the Polyakov loops can be calculated in any Berry gauge. Starting from the simplest $m_0=n_0=0$ case of a centered trap, in Appendix \ref{app:polya} we show how it is possible to explicitly construct a Berry gauge that is well defined and periodic around the MBZ along the entire $k_x=0$ line for all $k_y \in [0, \mathcal{L}_y]$ while, at the same time, satisfying the condition $\mathcal{A}_y(k_x=0,k_y) = 0$. Thanks to the periodicity of the Berry gauge, the transition function in this Berry gauge is then $\phi_y(k_x=0) = 0$, and thus
\begin{align}
\Phi_y (0) =  \phi_y(0) -\int_0^{\mathcal{L}_y} \mathcal{A}_y (0,k_y^\prime)dk_y^\prime=0. \label{phiykx0}
\end{align}
From (\ref{phixphiy}), this directly implies that $q\theta_y=0$ in our Berry gauge choice (\ref{ourberrygauge})~\footnote{Note that in all formulas only the combination $q\theta_y$ appears and not $\theta_y$ separately.}.
Similarly, another gauge can be found (see also Appendix \ref{app:polya}) that allows for an easy calculation of $\Phi_x(k_y=0)$, giving $\theta_x=0$.

As the position and the Berry connection are coupled as $\mathbf{r} + \mathcal{A}(\mathbf{p})$, for a generic trap position the shift of the harmonic trap center to $\mathbf{r}_0 = (m_0, n_0) \neq 0$ leads to a constant shift of the Berry connection by $-(m_0, n_0)$,
\begin{equation}
(\mathcal{A}_x, \mathcal{A}_y) = (0, \bar{\Omega}k_x + \delta \mathcal{A}_y) - (m_0, n_0).
\end{equation}
Regarding the MBZ as a two-dimensional torus embedded in a three-dimensional space, the effect of a shift of the trap position is to insert a Berry flux through the central hole and the body of the torus.
To bring this expression into the form (\ref{ourberrygauge}) of the Berry connection, a Berry gauge transformation can be performed which transfers the information on the trap position into the $\theta_{x,y}$ angles,
\begin{eqnarray}
\theta_x &=& 2\pi m_0/q, \\ 
\theta_y &=& 2\pi n_0/q.
\end{eqnarray}
This is a key result of this work: the angles $\theta_{x,y}$ parametrizing the momentum-space transition functions $\phi_{x,y}(k_{y,x})$ around the 2D torus can be experimentally varied by simply moving the center of the harmonic trap.

\section{Noninteracting Momentum-space Harper-Hofstadter model}
\label{sec:nonint}

With all these ingredients, we can now explicitly construct the Hamiltonian of the momentum-space HH model. Within the tight-binding model, we define $\alpha_{\mu,\nu}$ to be the annihilation operators of a Wannier state localized around $\mathbf{k}=(2\pi \mu/q, 2\pi \nu/q)$ in momentum space.
As a consequence of the toroidal topology, only states inside one MBZ are effectively independent, and the boundary conditions (\ref{momentumboundary}) and (\ref{boundaryexpression}) around the torus as $k_x \to k_x + 2\pi/q$ or $k_y \to k_y + 2\pi$ imply that
\begin{eqnarray}
	\alpha_{\mu+1,\nu} &=& e^{i\phi_x (2\pi \nu/q)}\alpha_{\mu,\nu}, \\
	\alpha_{\mu,\nu+q} &=& e^{i\phi_y (2\pi \mu/q)}\alpha_{\mu,\nu}.	
\end{eqnarray}
For our choice of Berry gauge (\ref{ourberrygauge}), the relevant values of the transition functions are 
\begin{align}
	\phi_x (2\pi \nu/q)
	&=
	\theta_x + \int_{0}^{\mathcal{L}_x} dk_x^\prime \int_0^{2\pi \nu/q}dk_y^\prime \Omega(k_x^\prime, k_y^\prime)
	\notag \\
	&=
	\theta_x + \frac{2\pi}{q} \frac{2\pi \nu}{q} \bar{\Omega}
	\notag \\
	&=
	\theta_x + 2\pi \mathcal{C} \nu /q
\end{align}
and $\phi_y=q\theta_y$.

Then, hopping along the $k_x$ direction from and to the MBZ has the form
\begin{align}
	-J^\prime &(\alpha_{1,\nu}^\dagger \alpha_{0,\nu} + \alpha_{0,\nu}^\dagger \alpha_{-1,\nu} + \mathrm{H.c.})/2
	\notag \\
	&=
	-2 J^\prime \cos (\theta_x + 2\pi \mathcal{C} \nu/q)\alpha_{0,\nu}^\dagger \alpha_{0,\nu},
\end{align} 
where $J^\prime > 0$ is the momentum-space hopping amplitude which we will estimate shortly.
Hopping along $k_y$ has an analogous form proportional to the same $J'$ and, with our choice for the Berry gauge (\ref{ourberrygauge}), does not involve any hopping phase.

Summing up all terms, we obtain the final Hamiltonian for the momentum-space HH model:
\begin{align}
	&\mathcal{H}^{\mathrm{M}}=-J^\prime
	\sum_{\nu = 0}^{q-1}
	\left(
	\alpha_{\nu+1}^\dagger \alpha_\nu + \alpha_\nu^\dagger \alpha_{\nu+1}
	\right.
	\notag \\
	&\hspace{3cm}
	\left.
	+ 2\cos (\theta_x + 2\pi \mathcal{C} \nu/q)\alpha_\nu^\dagger \alpha_\nu
	\right), \label{momentumhamiltonian}
\end{align}
where we have defined the tight-binding operators within the MBZ $\alpha_\nu \equiv \alpha_{0,\nu}$ and $\alpha_q \equiv e^{iq \theta_y}\alpha_0$.
We note that our Hamiltonian~(\ref{momentumhamiltonian}) is formally analogous to the momentum-space representation of the ordinary real-space HH model, but the roles of the momenta $k_{x,y}$ are replaced here by the angles $\theta_x$ and $\theta_y$. Further analysis of $\theta$-space topology, such as the effect of the Berry curvature and the Chern number in $\theta$ space, is discussed elsewhere~\cite{Future}.

To quantitatively validate the momentum-space HH model Hamiltonian (\ref{momentumhamiltonian}), we proceed with a specific example, $\alpha = 1/4$, which has direct experimental relevance~\cite{Aidelsburger2013,Aidelsburger2014,Hafezi2013, Miyake2013}.
The four eigenvalues and eigenvectors of the momentum-space HH Hamiltonian (\ref{momentumhamiltonian}) are straightforwardly obtained by diagonalizing a $4 \times 4$ matrix. In the simplest $m_0 = n_0=0$ ($\theta_x =\theta_y= 0$) case of a perfectly centered trap, for instance, the eigenvalues are $\pm 2\sqrt{2} J^\prime$ and 0 (doubly degenerate).

\begin{figure}[htbp]
\begin{center}
\includegraphics[width=7.0cm]{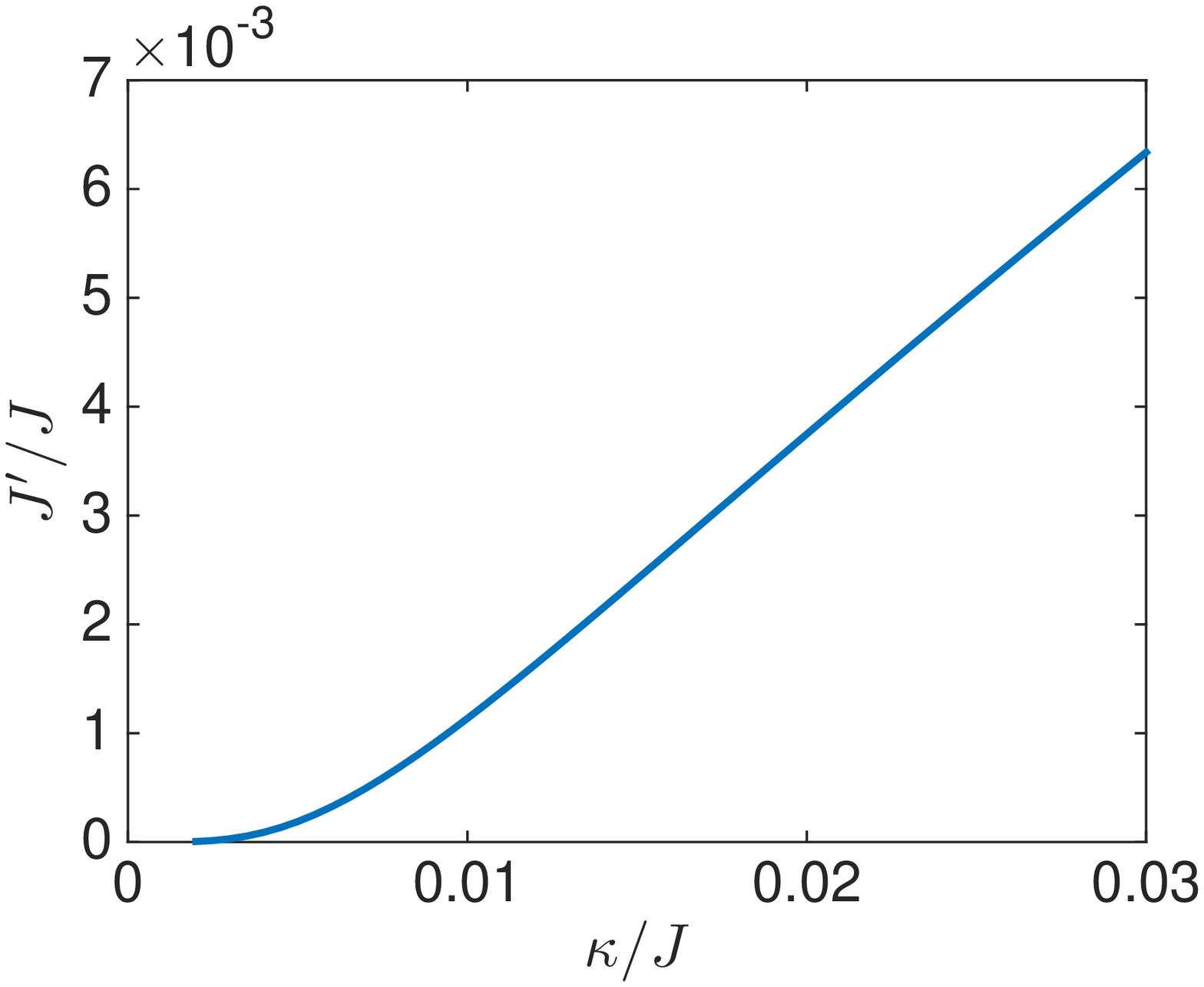}
\caption{Numerical estimate of $J^\prime$ as a function of $\kappa$ (both in units of $J$) for an $\alpha=1/4$ real-space HH model. Numerical calculation was performed on a $80\times80$ real-space lattice.}
\label{jprime}
\end{center}
\end{figure}

Using this spectrum, we can numerically estimate the magnitude of the momentum-space hopping $J^\prime$ as a function of the trap strength $\kappa$. 
Namely, the energies and eigenstates of the original real-space Hamiltonian (\ref{hamiltonian}) can be numerically obtained by simple diagonalization on a (sufficiently large) finite real-space lattice. By setting the numerically calculated energy gap between the two lowest states to be $2\sqrt{2}J^\prime$, we estimate $J^\prime$.
In Fig.~\ref{jprime}, we plot the estimated $J^\prime$ as a function of $\kappa$ for a range of $\kappa$ where the momentum-space tight-binding picture is appropriate (see below). Of course, an analogous calculation can be performed for any $\alpha$.

\begin{figure}[htbp]
\begin{center}
\includegraphics[width=6.0cm]{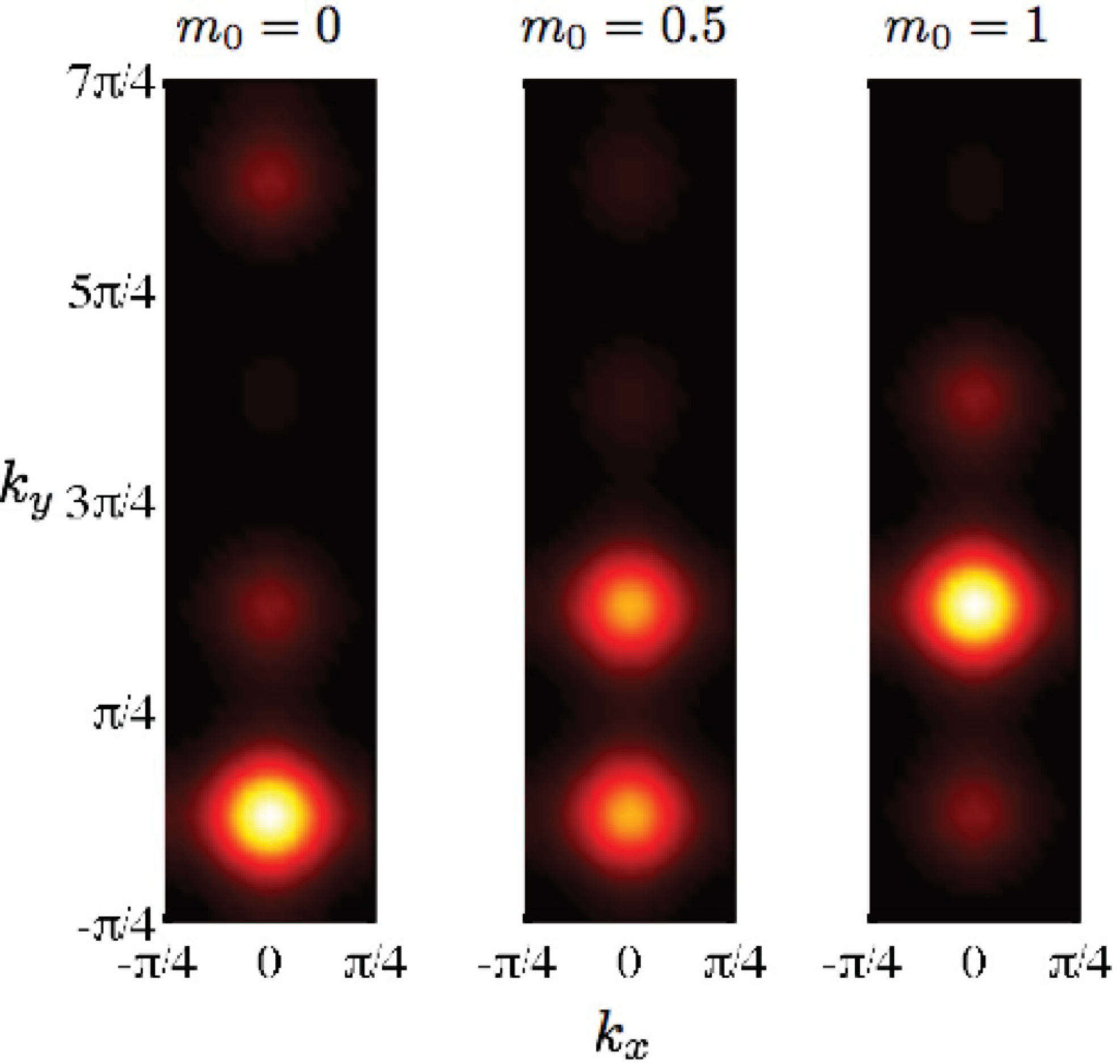}
\caption{The wave function of the $\alpha = 1/4$ ground state in the MBZ for $n_0=0$ and $m_0 = 0$, 0.5, and 1, from left to right, numerically calculated on an $80\times 80$ real-space lattice with $\kappa = 0.01J$.}
\label{momentumpsi}
\end{center}
\end{figure}

The numerical prediction for the momentum-space wave function in the MBZ is directly found by Fourier transforming the numerical wave function and then summing the square modulus of all states differing by reciprocal lattice vectors~\cite{Price2014}.
In Fig.~\ref{momentumpsi} we plot the square modulus of the momentum-space wave function summed over all bands in the MBZ for $n_0 = 0$ and three different values of $m_0$.

As a most remarkable feature, we see that as one varies $m_0$, the momentum-space wave function ``moves" in the $k_y$ direction. This can be understood in terms of Laughlin's {\it Gedankenexperiment} on the quantum Hall effect~\cite{Laughlin1981} now in momentum space; in the framework of fixing the boundary condition and varying the Berry connection, increasing $m_0$ corresponds to inserting magnetic flux in the momentum-space torus along the $k_x$ direction. The momentum-space wavefunction moves along the $k_y$ direction, coming back to the original state after changing $m_0$ by $q$, which corresponds to inserting one flux quantum in momentum space.

\begin{figure}[htbp]
\begin{center}
\subfigure[$\kappa = 0.005J$]{
\includegraphics[width=4.0cm]{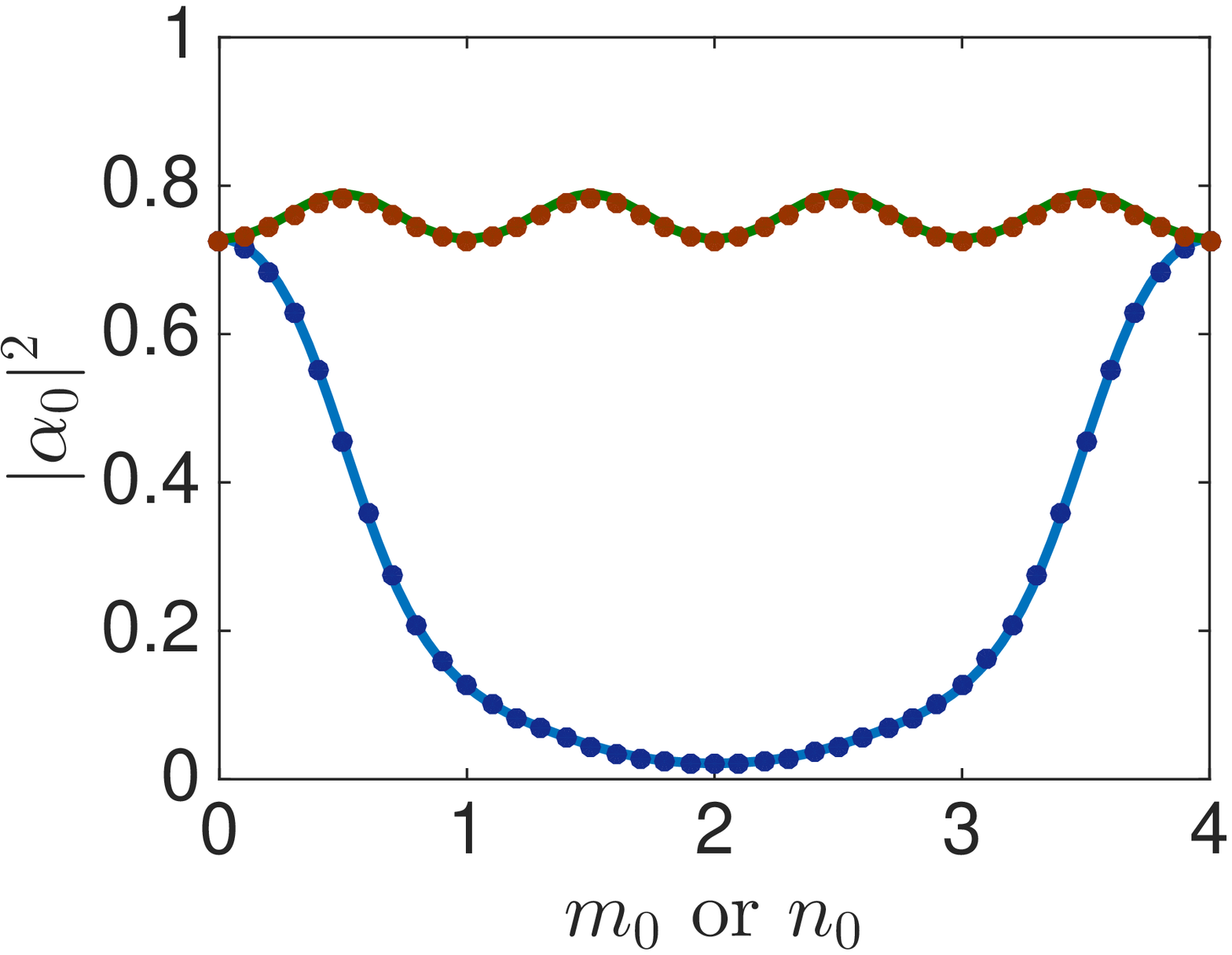}
}
\subfigure[$\kappa = 0.01J$]{
\includegraphics[width=4.0cm]{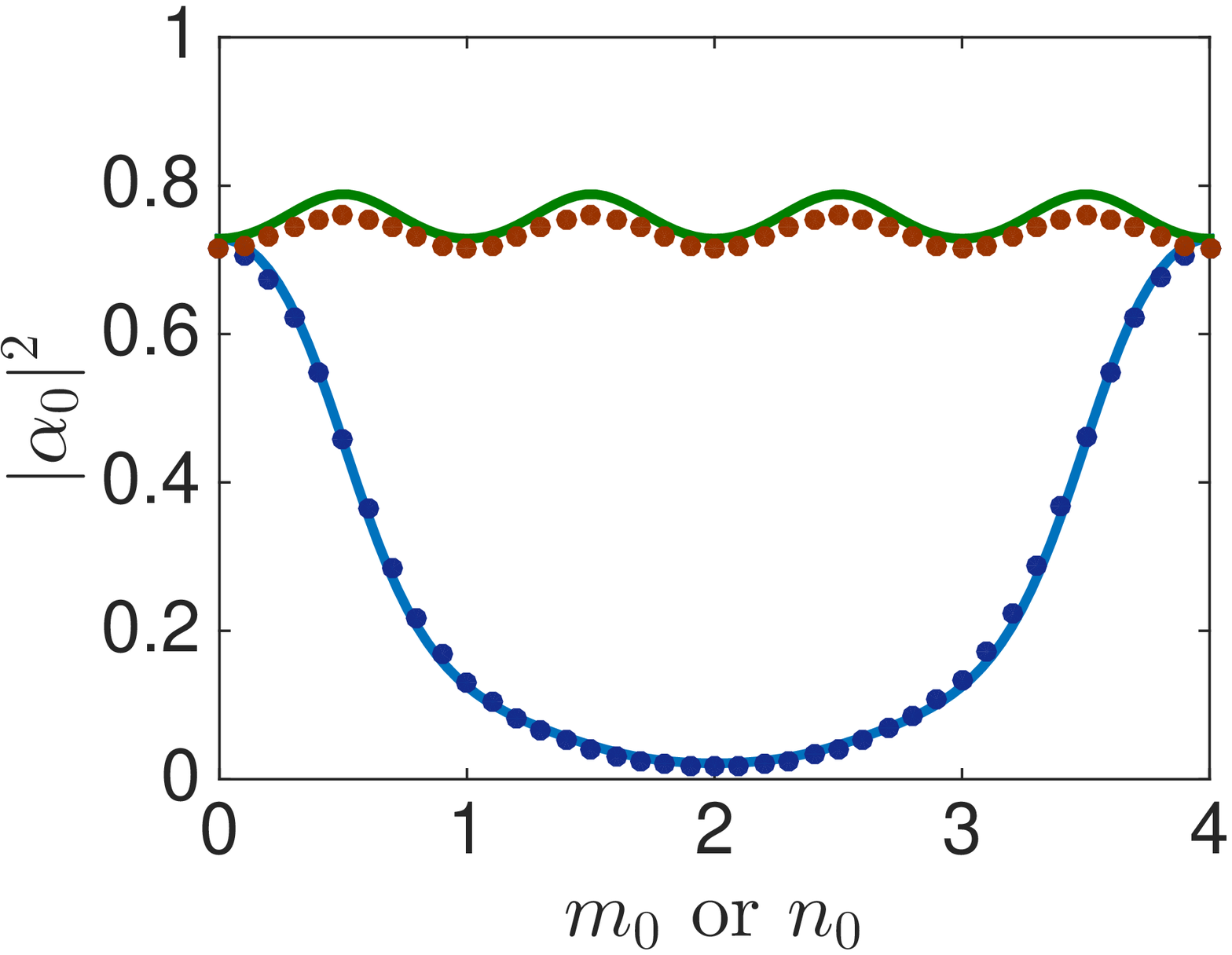}
}
\subfigure[$\kappa = 0.02J$]{
\includegraphics[width=4.0cm]{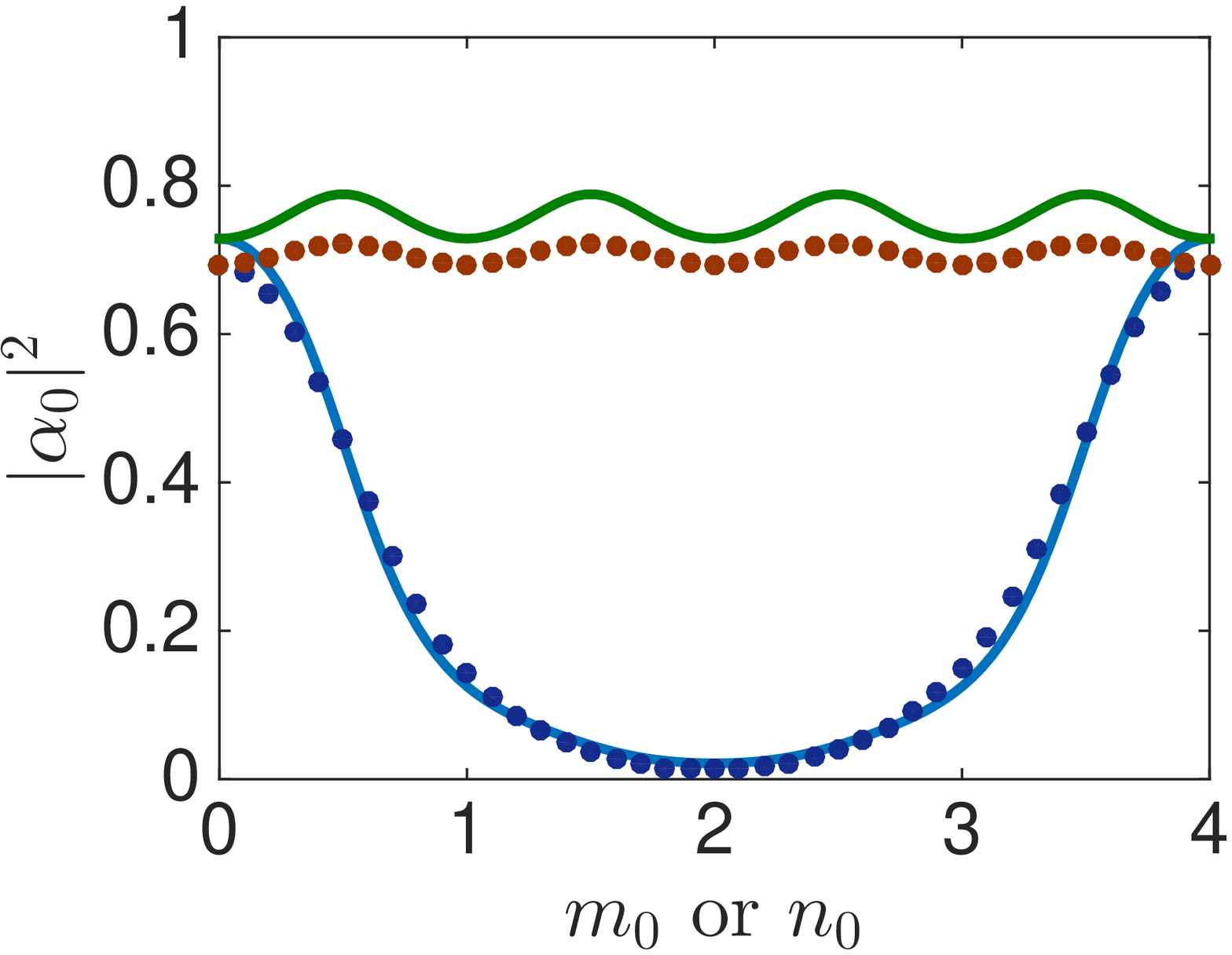}
}
\subfigure[$\kappa = 0.03J$]{
\includegraphics[width=4.0cm]{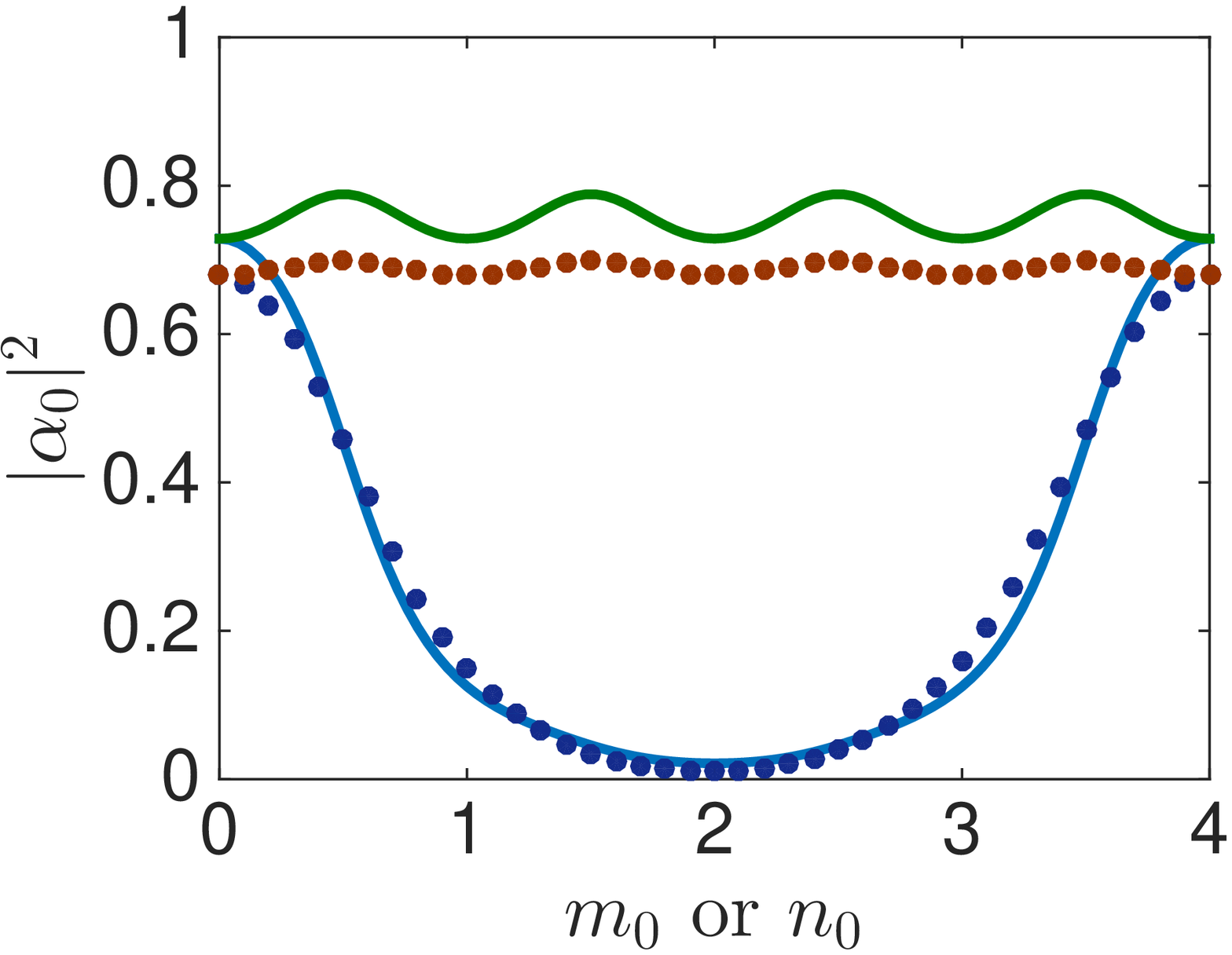}
}
\caption{The weight $|\alpha_0|^2$ as a function of $m_0$ for $n_0 = 0$ (lower line and dots) and as a function of $n_0$ for $m_0 = 0$ (upper line and dots) for four different values of $\kappa$: (a) $\kappa = 0.005J$, (b) $\kappa = 0.01J$, (c) $\kappa = 0.02J$, and (d) $\kappa = 0.03J$. The solid lines are the prediction from the momentum-space HH model. The dots are numerical results, calculated on a $80\times 80$ real-space lattice.}
\label{tightbinding}
\end{center}
\end{figure}

In order to compare the full numerical solution of the trapped real-space HH model (3) with the analytical momentum-space tight-binding model, the occupations $|\alpha_\nu|^2$ of the four momentum-space tight-binding states can be extracted from the numerics by integrating the square modulus of the wave function in the corresponding region: $\frac{\nu\pi}{2}-\frac{\pi}{4} \le k_y \le \frac{\nu\pi}{2}+\frac{\pi}{4}$. In Fig.~\ref{tightbinding}, the numerical estimate for $|\alpha_0|^2$ as a function of $m_0$ and $n_0$ is compared to the analytical tight-binding model for four different values of $\kappa$.
The agreement between the numerical and analytical results is particularly good for small values of $\kappa$. Qualitatively, this can be understood as a consequence of the smaller momentum-space kinetic energy, leading to a more accurate momentum-space tight-binding description. When the harmonic trap strength $\kappa$ increases, the momentum-space kinetic energy becomes larger. Then, the contribution from higher-energy modes in the momentum-space lattice becomes more and more important, and the single-band momentum-space tight-binding approximation is violated.

More quantitatively, the regime of validity of the momentum-space HH model can be estimated by comparing the bandwidth $\Delta E_{\mathrm{band}}$ to the ``recoil energy" $E_r = (2\pi/\lambda)^2 / 2m$, with $m$ and $\lambda/2$ being the mass and the lattice spacing, respectively.
In the sinusoidal optical lattice, it is known that when the lattice potential $V_0$ is sufficiently deep such that $V_0/E_r \simge 5$~\cite{PitaevskiiStringari, BlochBook}, the tight-binding approximation is valid. Applying this criterion to the momentum-space HH model with $\alpha = 1/4$, where the lattice depth is given by the bandwidth $\Delta E_{\mathrm{band}} \sim 0.22J$, the lattice spacing in momentum space is $\pi/2$, and the effective mass is $1/\kappa$, one finds the tight-binding condition 
\begin{equation}
\kappa \simle \Delta E_{\mathrm{band}}/10 \simeq 0.02\,J. 
\label{TBbound}
\end{equation}
In Fig.~\ref{tightbinding}, we can observe that the deviation from the model indeed starts to become significant for $\kappa > 0.02J$, while its predictions become more and more exact as $\kappa$ is further reduced.
The deviation between the momentum-space HH prediction and the full numerical simulation at $\kappa > 0.02J$ can also be seen later in Fig.~{\ref{phasediagram}} for the interacting ground states.

For decreasing values of $\alpha$, the band dispersion becomes flatter and flatter with an exponentially decreasing $\Delta E_\mathrm{band}\to 0$. This imposes more and more stringent bounds (\ref{TBbound}) so that our tight-binding model is accurate for only an exponentially weak trap strength $\kappa$. In this regime of a very flat band dispersion $\Delta E_{\mathrm band}\to 0$, which is realized for $\alpha\ll 1$, the momentum-space wave function is delocalized over the whole MBZ, and the eigenstates reduce to the toroidal Landau levels studied in~\cite{Price2014}.

Experimentally, time-of-flight measurements reveal the (physical) momentum distribution of the condensate. To obtain a quasimomentum distribution similar to that in Fig.~\ref{momentumpsi}, one then needs to sum the square modulus of the wave function of states differing by reciprocal lattice vectors in a way analogous to how we numerically obtained Fig.~\ref{momentumpsi} from the real-space wave function of the full model (\ref{hamiltonian}). We note that both the physical momentum and quasimomentum distributions are magnetic gauge dependent and hence depend on the specific gauge realized in experiments~\cite{Kennedy2015}.

\section{Interacting System}

Now we consider the effect of on-site interactions, described by the Hamiltonian
\begin{align}
	\mathcal{H}_{\mathrm{int}}
	=
	\frac{U}{2}\sum_{m,n} a_{m,n}^\dagger a_{m,n}^\dagger a_{m,n} a_{m,n}. \label{hint}
\end{align}
Assuming that the average number of particles per site is large enough, we employ the usual Bogoliubov mean-field theory, replacing the $a_{m,n}$ operators by $\mathbf{C}$ numbers. The number of particles $N=\sum_{m,n}|a_{m,n}|^2$.
The ground state in the homogeneous $\kappa=0$ case without a trap was found in~\cite{Powell2010, Powell2011} to exhibit a nontrivial spatial order.

\subsection{Numerical ground states}

As a first step to understanding the trapped case, we use the imaginary-time propagation method to find the ground state of $\mathcal{H}_0 + \mathcal{H}_{\mathrm{int}}$ for $\alpha = 1/4$ and a trap center  at $(m_0, n_0) = (0,0)$, which preserves the rotational symmetry of the underlying lattice.
We note that since the Harper-Hofstadter Hamiltonian breaks ordinary rotational symmetry due to the complex hopping, the correct (generalized) $\pi/2$ rotational symmetry $\mathcal{R}$ is a combination of a rotation and an appropriate phase change, which in the Landau magnetic gauge reads~\cite{Balents2005,Powell2010,Powell2011}
\begin{align}
	\mathcal{R} a_{m,n} \mathcal{R}^\dagger = a_{-n,m}e^{i2\pi \alpha mn}. \label{rotation}
\end{align}
Application of $\mathcal{R}$ four times over transforms the operator into itself.
The Hamiltonian is invariant under this generalized rotational symmetry.

For a given value of the trap strength $\kappa$, we observe three structural changes in the ground state wave function, from rotationally symmetric to nonsymmetric states as one increases $UN$. Even though our calculations have been performed in a Landau form (\ref{A_magn}) of the magnetic gauge, the change in the structure of the ground state is a magnetic-gauge-independent feature that does not depend on the specific choice of the vector potential $A_{x,y}$ and the real-space hopping phases in (\ref{hamiltonian}).

\begin{figure}[htbp]
\begin{center}
\subfigure[$UN = 0.04J$, $Z_4$]{
\includegraphics[width=2.8cm]{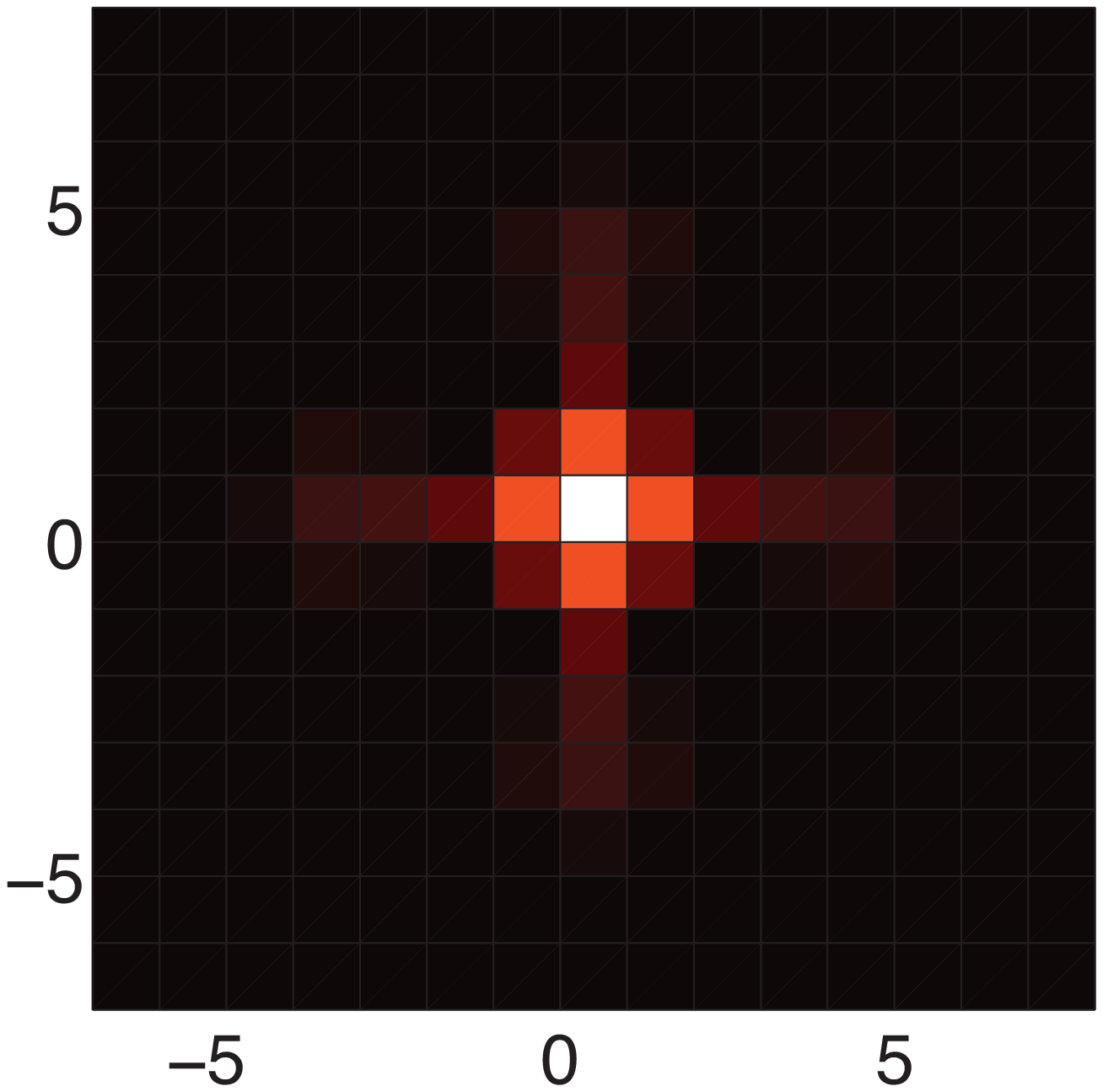}
}
\subfigure[$UN = 0.11J$, $Z_2$]{
\includegraphics[width=2.8cm]{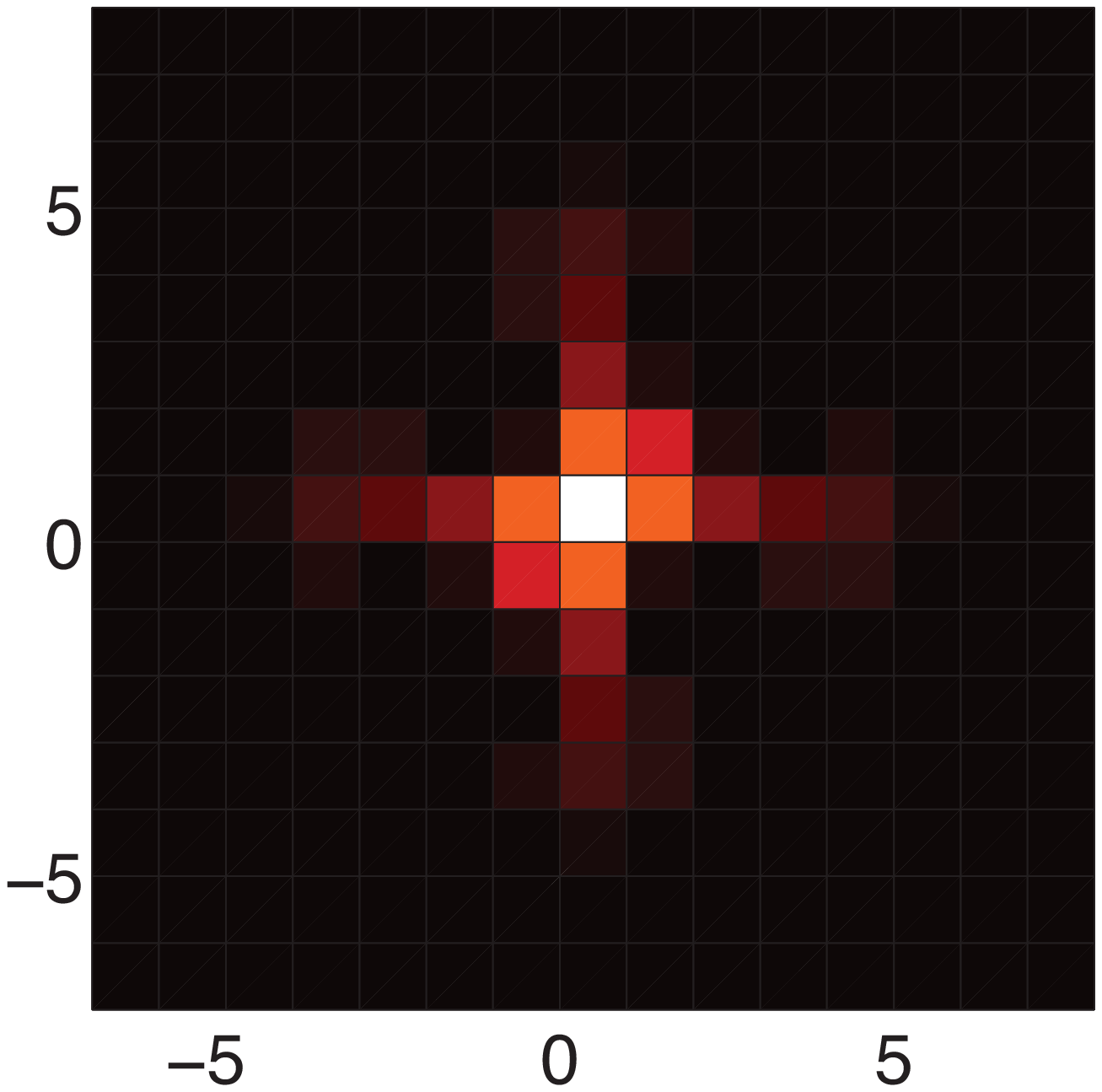}
}
\subfigure[$UN = 0.14J$, $Z_\times$]{
\includegraphics[width=2.8cm]{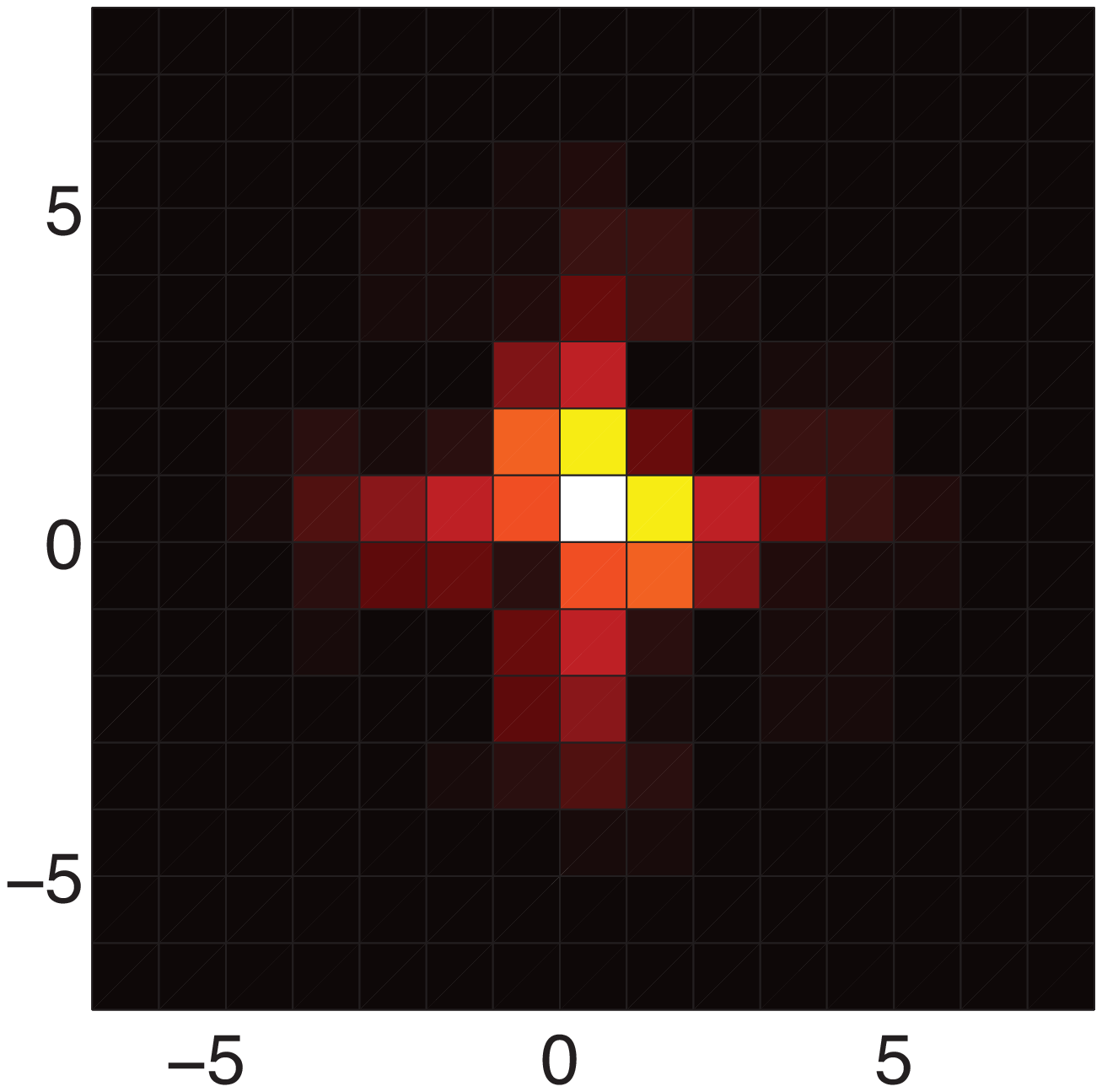}
}
\subfigure[$UN = 0.40J$, $Z_2$]{
\includegraphics[width=2.8cm]{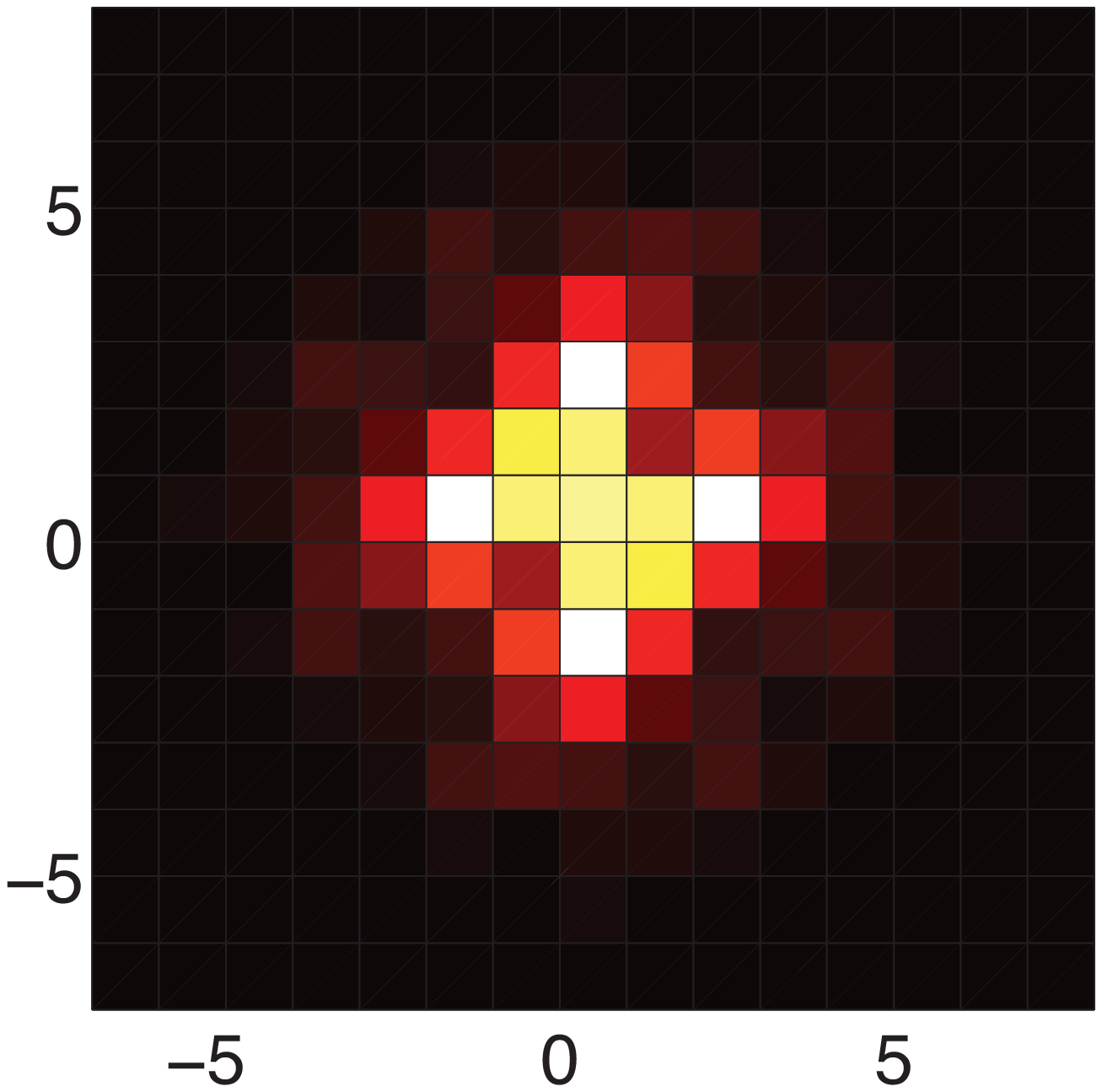}
}
\caption{Density profile $|a_{m,n}|^2$ of the ground-state wave function numerically obtained by the imaginary-time propagation method on a $40\times40$ real-space lattice with $\kappa = 0.01J$ for different values of $UN$: (a) $UN = 0.04J$, $Z_4$ state, (b) $UN = 0.11J$, $Z_2$ state, (c) $UN = 0.14J$, $Z_\times$ state, and (d) $UN = 0.40J$, $Z_2$ state. Only the central region is plotted.}
\label{groundpsi}
\end{center}
\end{figure}

The full series of structural changes is illustrated in the different panels of Fig.~\ref{groundpsi}, where we plot the ground-state wave function for different values of $UN$.
When $UN$ is small, the ground state has the full $90^\circ$ rotational symmetry ($Z_4$ state). As one increases $UN$, the ground state changes first to a state with only $180^\circ$ rotational symmetry and double degeneracy ($Z_2$ state) and then to a state with no rotational symmetry and fourfold degeneracy ($Z_\times$ state). As one further increases $UN$, the ground state recovers its $Z_2$ character.
Later in Fig.~\ref{phasediagram}, the changes of ground states $Z_4 \to Z_2$, $Z_2 \to Z_\times$, and $Z_\times \to Z_2$ are indicated by dots.
As expected, the degeneracy of the ground states in the $Z_2$ and $Z_\times$ cases is lifted by any spatial displacement of the trap center which explicitly breaks the rotational or reflection symmetry of the Hamiltonian (\ref{hamiltonian}).

\subsection{Momentum-space HH model with interaction}

We now investigate the symmetry breaking in the ground state using the momentum-space HH model.
To study the effects of interaction in momentum space, we write $\mathcal{H}_{\mathrm{int}}$ in terms of momentum-space operators and then project onto the lowest band and retain operators only at the dispersion minima $(0,\pi \nu/2)$, where $\nu = 0,1,2,$ or $3$. 
For our chosen Landau form for the Berry gauge (\ref{ourberrygauge}), with a vanishing Berry connection along $k_x = 0$,
the interaction Hamiltonian is (see Appendix \ref{app:inter} for details)
\begin{align}
	\mathcal{H}_{\mathrm{int}}^{\mathrm{M}}
	=
	\frac{U^\prime}{16}
	\left(
	9 I_1 + I_2 - 6 I_3 + 4 I_4
	\right), \label{hamint}
\end{align}
where $U^\prime$ is a constant and
\begin{align}
	I_1
	&\equiv
	\left(|\alpha_0|^2 + |\alpha_1|^2 + |\alpha_2|^2 + |\alpha_3|^2 \right)^2,
	\notag \\
	I_2
	&\equiv
	(
	\alpha_0^\dagger \alpha_2 + \alpha_2^\dagger \alpha_0 - \alpha_1^\dagger \alpha_3 - \alpha_3^\dagger \alpha_1
	)^2,
	\notag \\
	I_3
	&\equiv
	|\alpha_0|^2 |\alpha_1|^2 + |\alpha_1|^2 |\alpha_2|^2 + |\alpha_2|^2 |\alpha_3|^2 + |\alpha_3|^2 |\alpha_0|^2
	\notag \\
	&\phantom{-}-\alpha_0^\dagger\alpha_1^\dagger\alpha_2\alpha_3
	-\alpha_1^\dagger\alpha_2^\dagger\alpha_3\alpha_0
	-\alpha_2^\dagger\alpha_3^\dagger\alpha_0\alpha_1
	-\alpha_3^\dagger\alpha_0^\dagger\alpha_1\alpha_2,
	\notag \\
	I_4
	&\equiv
	\alpha_0^\dagger \alpha_2^\dagger \alpha_1^2 + \alpha_0^\dagger \alpha_2^\dagger \alpha_3^2
	+ \alpha_1^\dagger \alpha_3^\dagger \alpha_0^2 + \alpha_1^\dagger \alpha_3^\dagger \alpha_2^2
	+ \mathrm{H.c.}
	\notag \\
	&\phantom{-}
	-4|\alpha_0|^2|\alpha_2|^2 - 4|\alpha_1|^2 |\alpha_3|^2
\end{align}
are the four-operator combinations of momentum-space tight-binding field operators that are invariant under the symmetry of the HH model with $\alpha = 1/4$~\cite{Balents2005}.

To find the ground state of the momentum-space HH model, we minimize $\mathcal{H}^{\mathrm{M}} + \mathcal{H}_{\mathrm{int}}^{\mathrm{M}}$, treating $\alpha_\nu$ as independent complex variables with the constraint $\sum_{\nu=0}^3 |\alpha_\nu|^2 = N$.
Since the numerical simulation in Fig.~\ref{groundpsi} shows symmetry-breaking changes of the ground state, it is convenient to work in the basis of rotational eigenstates in order to find where the rotational-symmetry breaking occurs.
For our chosen Berry gauge (\ref{ourberrygauge}), for which the Berry connection is zero along the $k_x = 0$ line, one can show that momentum-space variables $\alpha_\nu$ transform under the generalized rotation $\mathcal{R}$ in (\ref{rotation}) as
\begin{align}
	\mathcal{R} \alpha_\nu \mathcal{R}^\dagger = \frac{1}{\sqrt{q}}\sum_{\nu^\prime} e^{i2\pi \alpha \nu \nu^\prime}\alpha_{\nu^\prime}.
\end{align}
This holds for any value of the magnetic flux $\alpha = p/q$.
Setting $\alpha = 1/4$, the rotation of $\alpha_\nu$ is represented by a matrix
\begin{align}
	\mathcal{R}
	=
	\frac{1}{2}
	\begin{pmatrix}
	1 & 1 & 1 & 1 \\
	1 & i & -1 & -i \\
	1 & -1 & 1 & -1 \\
	1 & -i & -1 & i \\
	\end{pmatrix}.
\end{align}
The eigenvalues of this matrix are $1$ (doubly degenerate), $i$, and $-1$, and the corresponding eigenvectors can be taken as
\begin{align}
	\mathbf{v}_0
	&=
	\frac{1}{2\sqrt{2}}
	\begin{pmatrix}
	1+\sqrt{2} \\ 1 \\ -1+\sqrt{2} \\ 1
	\end{pmatrix},
	&
	\mathbf{v}_1
	&=
	\frac{1}{2\sqrt{2}}
	\begin{pmatrix}
	1-\sqrt{2} \\ 1 \\ -1-\sqrt{2} \\ 1
	\end{pmatrix},
	\notag \\
	\mathbf{v}_2
	&=
	\frac{1}{\sqrt{2}}
	\begin{pmatrix}
	0 \\ 1 \\ 0 \\ -1
	\end{pmatrix},
	&
	\mathbf{v}_3
	&=
	\frac{1}{2}
	\begin{pmatrix}
	1 \\ -1 \\ -1 \\ -1
	\end{pmatrix},
\end{align}
and then we define the rotational eigenstates $\beta_\nu$ by
\begin{align}
	\beta_\nu \equiv \mathbf{v}_\nu^\dagger \cdot
	\begin{pmatrix}
	\alpha_0 \\ \alpha_1 \\ \alpha_2 \\ \alpha_3
	\end{pmatrix}. \label{betadef}
\end{align}
Following from the eigenvalues of $\mathcal{R}$ mentioned above, $\beta_\nu$ transforms under a real-space rotation as $\beta_0 \to \beta_0$, $\beta_1 \to \beta_1$, $\beta_2 \to i\beta_2$, and $\beta_3 \to -\beta_3$.

Using (\ref{betadef}), one can write the momentum-space Harper-Hofstadter Hamiltonian $\mathcal{H}^{\mathrm{M}} + \mathcal{H}_{\mathrm{int}}^{\mathrm{M}}$ in terms of $\beta_\nu$. Also in the new variables, the interaction term keeps a complicated form analogous to (\ref{hamint}).

\begin{figure}[htbp]
\begin{center}
\includegraphics[width=7.5cm]{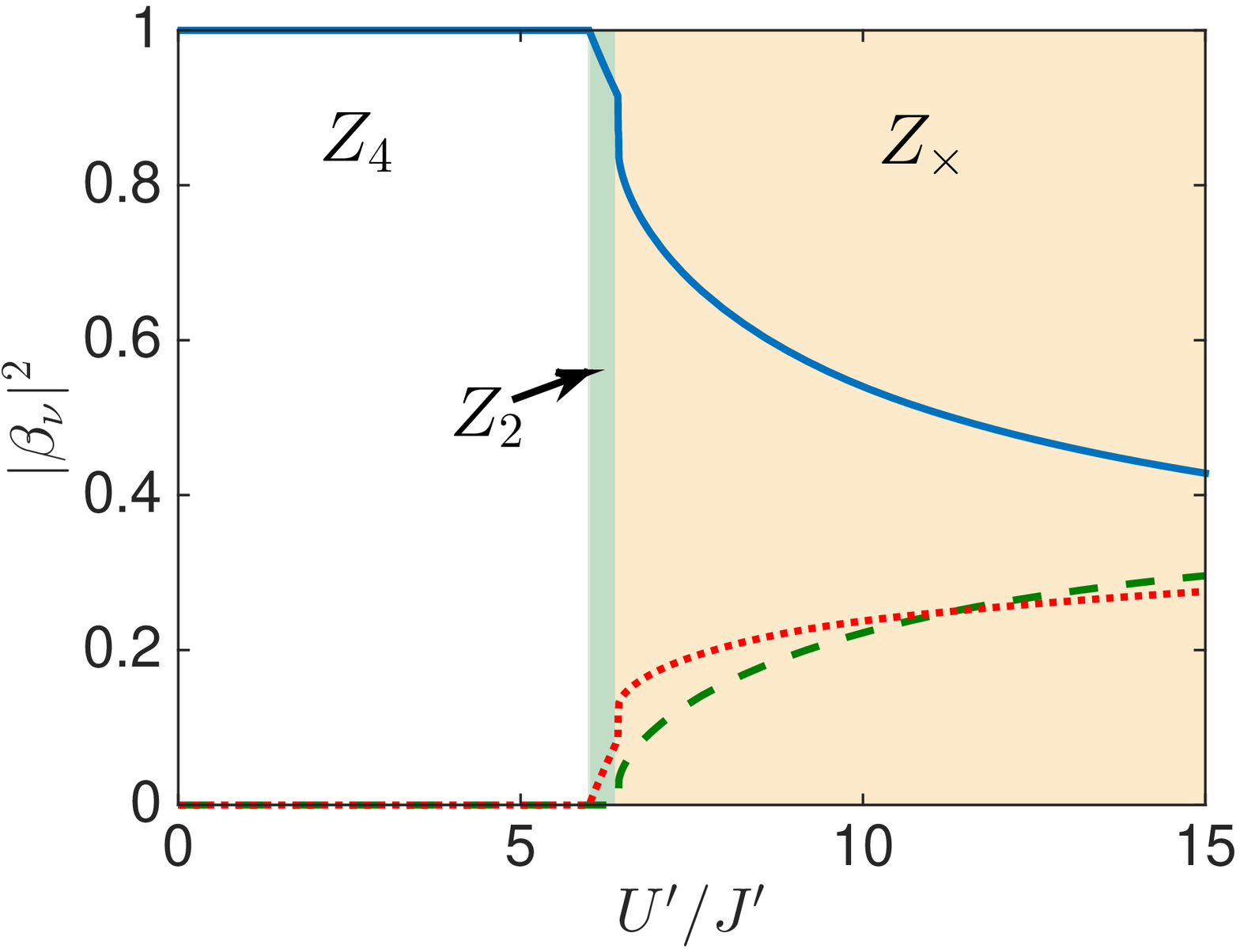}
\caption{The weights of rotational eigenstates $|\beta_0|^2 + |\beta_1|^2$ (blue solid line), $|\beta_2|^2$ (green dashed line), and $|\beta_3|^2$ (red dotted line) as a function of $U^\prime/J^\prime$, obtained by minimizing the momentum-space Harper-Hofstadter Hamiltonian.}
\label{phasetransition}
\end{center}
\end{figure}

The ground state is obtained by finding the states which minimize $\mathcal{H}^{\mathrm{M}} + \mathcal{H}_{\mathrm{int}}^{\mathrm{M}}$ under the constraint $|\beta_0|^2 + |\beta_1|^2 + |\beta_2|^2 + |\beta_3|^2 = 1$.
In Fig.~\ref{phasetransition}, we plot the values of $|\beta_0|^2 + |\beta_1|^2$, $|\beta_2|^2$, and $|\beta_3|^2$ for the field configuration which minimizes $\mathcal{H}^{\mathrm{M}} + \mathcal{H}_{\mathrm{int}}^{\mathrm{M}}$ as a function of $U^\prime/J^\prime$.
We find that there is a change in ground-state symmetry at $U^\prime/J^\prime = 6$, above which different rotational eigenstates mix and the rotational symmetry is broken.
In the region $0 < U^\prime/J^\prime < 6.0$, the full rotational symmetry is present ($Z_4$ state).
In the region $6< U^\prime/J^\prime < 6.4$, only $\beta_2$ is zero, and the ground state breaks $\mathcal{R}$ symmetry but does not break $\mathcal{R}^2$ symmetry ($Z_2$ state).
In the region $6.4 < U^\prime/J^\prime$, no rotational symmetry is present ($Z_\times$ state).

\subsection{Comparison}

In order to make a quantitative comparison between the numerical simulation obtained from the imaginary-time propagation method and the results of the momentum-space HH model, we need to relate the $J^\prime$ and $U^\prime$ parameters of the momentum-space HH model to the parameters $\kappa$ and $UN$ of the real-space Hamiltonian (\ref{hamiltonian}). As already discussed in Sec.~\ref{sec:nonint} and illustrated in Fig.~\ref{jprime}, the value of $J^\prime$ is estimated from the numerical prediction for the energy gap between the two lowest states of the real-space Hamiltonian in the absence of interactions. Analogously, the value of $U^\prime$ is extracted by setting the numerically obtained interaction energy calculated from the noninteracting ground state equal to that of the analytical value of the momentum-space HH model, $35U^\prime/64$.

\begin{figure}[htbp]
\begin{center}
\includegraphics[width=7.0cm]{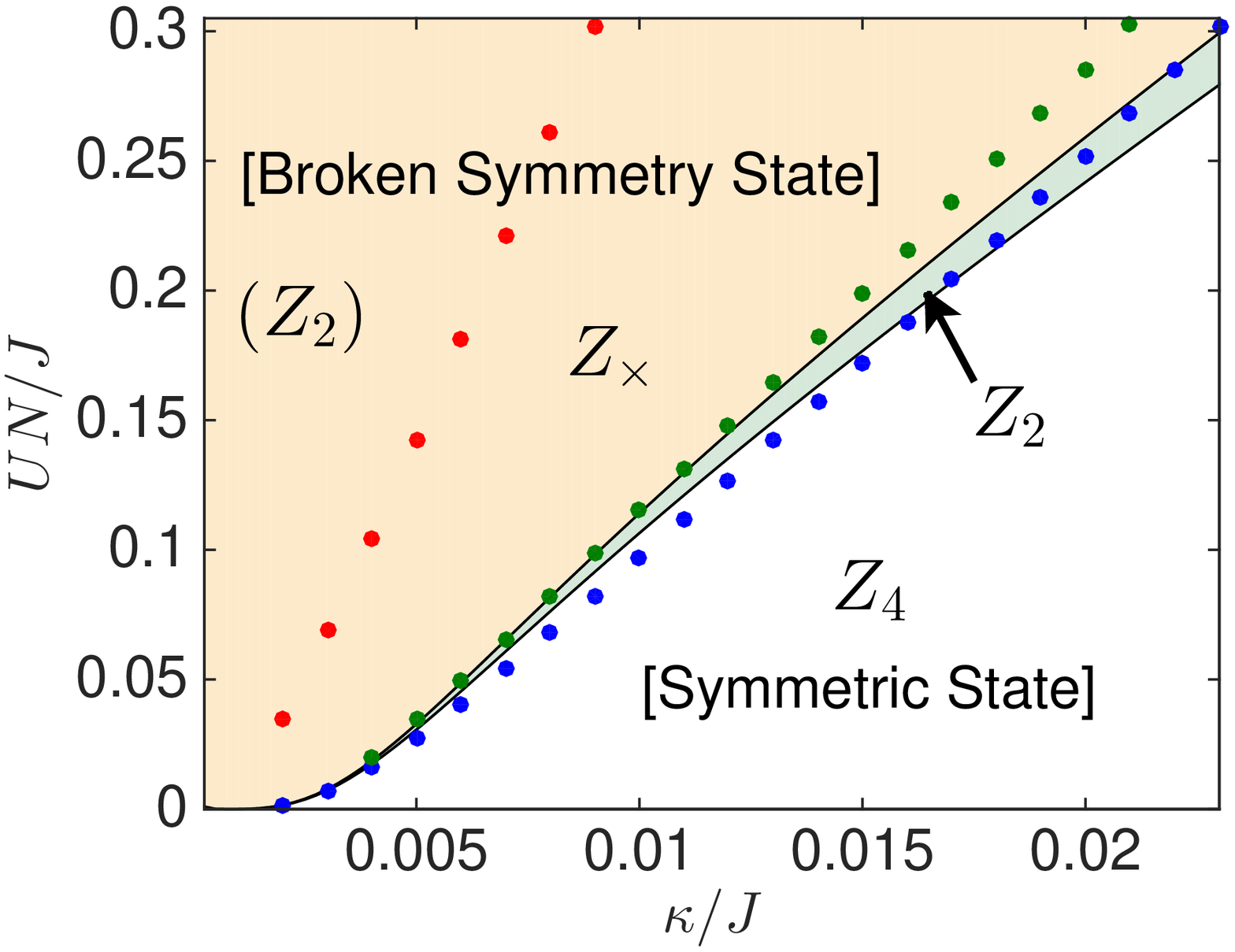}
\caption{Diagram showing how the symmetry of the ground state of the trapped interacting HH model changes as a function of the trap and the interaction strengths, with both axes in units of $J$. The solid lines are the predictions of the momentum-space HH model for the transition lines, while the dots show the corresponding predictions from numerical simulations performed via the imaginary-time propagation method on a $40 \times 40$ lattice.}
\label{phasediagram}
\end{center}
\end{figure}

In Fig.~\ref{phasediagram}, we show how the symmetry of ground states of the momentum-space HH model changes as a function of the parameters of the underlying real-space HH model.
For small $\kappa$, the two analytically predicted transition lines at $U^\prime/J^\prime = 6$ and $U^\prime/J^\prime = 6.4$ (solid lines) agree well with the numerical ones (dots), while deviations appear for larger $\kappa$ where the momentum-space single-band tight-binding approximation starts to fail. The situation is different for the last transition, $Z_\times \to Z_2$, which is not predicted by the momentum-space HH model. This absence is, however, fully compatible with the fact that for larger values of $UN$, the energy difference between different local energy minima decreases, so even a small contribution from upper bands beyond the tight-binding limit strongly affects the ground state.

Even though we have focused our attention here on the $\alpha=1/4$ case, we have ensured that the spontaneous breaking of rotational symmetry is a much more general feature of the momentum-space HH model. In particular, we have explicitly checked that similar symmetry-breaking transitions occur for $\alpha = 1/3$. It is, however, crucial to note that the main ingredient underlying the spontaneous breaking of rotation symmetry is the nonlocality of the interaction in momentum space: for a hypothetical purely local interaction in momentum space, the ground state of the momentum-space HH model on the torus would, in fact, be nondegenerate, and no change of the mean-field ground state would take place~\footnote{This statement may seem to be in contrast to the results of~\cite{Powell2011} that pointed out the possibility of spontaneous breaking rotational symmetries also in the presence of {\em local} interactions. However, this latter result was obtained for spatially extended lattices and no longer holds for small lattices like in our momentum-space HH model.}.

\section{Conclusion}
In this paper, we have shown how one can map a real-space Harper-Hofstadter model in the presence of a harmonic confinement onto a HH model in momentum space with nonlocal interactions. By shifting the trap center, we see an analog of Laughlin's charge-pumping {\em Gedankenexperiment} in momentum space. For sufficiently strong interactions, the nonlocal nature of the interactions in the momentum-space HH model is responsible for the appearance of degenerate ground states that spontaneously break rotational symmetry of the lattice.

Even though our discussion is focused on condensation in a real-space HH model, its conclusions extend to a variety of systems in which the energy dispersion of the bands shows multiple degenerate minima. In the longer run, this topological model raises a number of intriguing questions such as the possibility of observing quantum Hall effects in momentum space and creating new phases of matter under the effect of nonlocal interactions. 

\begin{acknowledgements}
This work was funded by ERC through the QGBE grant, by the EU-FET Proactive grant AQuS, Project No. 640800, and by Provincia Autonoma di Trento, partially through the project ``On silicon chip quantum optics for quantum computing and secure communications - SiQuro."
We thank Monika Aidelsburger for useful exchanges on their experiments and Nigel Cooper for stimulating discussions.
\end{acknowledgements}

\appendix

\begin{widetext}
\section{Evaluation of the Polyakov loops}
\label{app:polya}

In this appendix, we give more details on the calculation of the Polyakov loops for the real-space HH model in the $(m_0, n_0) = 0$ case.
We first discuss how to determine $\Phi_y$ and $\theta_y$.
We define the Fourier transform of the operators $a_{m,n}$ as
\begin{align}
	a_{m,n}
	=
	\sum_{\mathbf{k}\in \mathrm{MBZ}} e^{ik_x m + ik_y n} c_{m^{\prime \prime}}(\mathbf{k}), \label{fourier2}
\end{align}
where $m = qm^\prime + m^{\prime \prime}$, with $m^\prime$ being an integer and $m^{\prime \prime} = 0, 1, \cdots, q-1$.
The sum over the wave vectors is restricted to the first magnetic Brillouin zone.
In this basis, the momentum-space Hamiltonian is
\begin{align}
	\mathcal{H}_\mathbf{k}
	=
	-J
	\begin{pmatrix}
	2\cos k_y & e^{ik_x} & 0 & \cdots & e^{-ik_x}\\
	e^{-ik_x} & 2\cos (k_y - 2\pi \alpha) & e^{ik_x} & \cdots \\
	\vdots & \vdots & \vdots& \ddots & 0 \\
	e^{ik_x} & 0 & 0 & \cdots & 2\cos (k_y - (q-1)2\pi \alpha)
	\end{pmatrix}. \label{hamk2}
\end{align}
Then, when $k_x = 0$, the momentum-space Hamiltonian is a real symmetric matrix, which means that we can diagonalize the matrix by an orthogonal matrix.
Then, the transformation to the band basis can be done by a real matrix, and thus the Berry connection along $k_y$ is zero [$\mathcal{A}_y (0, k_y) = 0$].
The orthogonal matrix can also be taken as being smooth as one changes $k_y$ and periodic as $k_y \to k_y + \mathcal{L}_y$, which implies that the wave function along $k_x = 0$ is fully periodic in MBZ and the transition function associated with this Berry gauge choice is zero [$\phi_y (0, k_y) = 0$].
Then, the Polyakov loop in the $k_y$ direction is zero, $\Phi_y (0) = 0$.
On the other hand, in our chosen gauge (\ref{ourberrygauge}), $\Phi_y (0) = q \theta_y$, which then gives $q\theta_y = 0$.

Next, we discuss how to evaluate ${\Phi}_x(0)$ to determine $\theta_x$. Here we use a different convention for the Fourier transformation, corresponding to a different choice of the Berry gauge.
Instead of (\ref{fourier2}), we can choose to transform by
\begin{align}
	a_{m,n}
	=
	\sum_{\mathbf{k}\in \mathrm{BZ}}e^{ik_x m + ik_y n} \tilde{c}_\mathbf{k}, \label{fourier1}
\end{align}
where the sum is over the first full (not magnetic) Brillouin zone ($\mathrm{BZ}$). Then, we obtain the following Hamiltonian in momentum space:
\begin{align}
	\tilde{\mathcal{H}}_\mathbf{k}
	=
	-J
	\begin{pmatrix}
	2\cos k_x & e^{ik_y} & 0 & \cdots & e^{-ik_y}\\
	e^{-ik_y} & 2\cos (k_x + 2\pi \alpha) & e^{ik_y} & \cdots & \vdots \\
	\vdots & \vdots & \vdots& \ddots & 0 \\
	e^{ik_y} & 0 & 0 & \cdots & 2\cos (k_x + (q-1)2\pi \alpha)
	\end{pmatrix}, \label{hamk1}
\end{align}
where the basis is $\tilde{c}_{k_x,k_y}$, $\tilde{c}_{k_x + 2\pi \alpha,k_y}$, $\cdots$, $\tilde{c}_{k_x + (q-1) 2\pi \alpha ,k_y}$ and the overall factor $J>0$.
Then, when $k_y = 0$, the momentum-space Hamiltonian is a real symmetric matrix, and thus the $x$ component $\mathcal{A}_x$ of the Berry connection along the $k_y = 0$ line is zero for all $k_x$.

Under a wave-vector shift $k_x \to k_x + \mathcal{L}_x$ with $\mathcal{L}_x=2\pi/q$, the spectrum of $\tilde{\mathcal{H}}_\mathbf{k}$ is unchanged, while the eigenvectors experience a shift in their components, $(w_1, w_2, \cdots, w_{q-1},w_{q}) \to (w_2, \cdots, w_{q-1},w_{q},w_1)$. Since $\tilde{\mathcal{H}}_\mathbf{k}$ is an irreducible matrix with all the off-diagonal elements being nonpositive, the Perron-Frobenius theorem (see, e.g.,~\cite{Tasaki}) guarantees that for all $k_x$ the components of the smallest eigenvector can be taken as all positive. Therefore under a smooth variation $k_x \to k_x + \mathcal{L}_x$ the eigenvector is smoothly scanned $(w_1, w_2, \cdots, w_{q-1},w_{q}) \to (w_2, \cdots, w_{q-1},w_{q},w_1)$ without acquiring any extra factor. As a result, the transition phase in the considered gauge is $\phi_x(k_y=0)=0$. Using the definition (\ref{Poly_y}), it is straightforward to see that the Polyakov loop is also $\Phi_x(0)=0$. Thanks to the gauge independence of $\Phi_x$, this immediately implies that in our original Berry gauge choice (\ref{ourberrygauge}) one has $\theta_x=0$.

\section{Derivation of the interaction Hamiltonian in momentum space}
\label{app:inter}
Here we sketch the derivation of the interaction Hamiltonian of the momentum-space Harper-Hofstadter model.
We first expand the annihilation operator in the Fourier series according to (\ref{fourier2}) because, as noted above, with this Fourier transformation convention, the Berry connection along the $k_x = 0$ line can be taken to be zero, $\mathcal{A}_y(0,k_y)=0$, which is in accord with the Landau gauge we want to choose.
Then, the interaction in momentum space reads
\begin{align}
	\mathcal{H}_{\mathrm{int}}
	\propto
	\frac{U}{2}
	\sum_{\mathbf{k}_1 + \mathbf{k}_2 = \mathbf{k}_3 + \mathbf{k}_4}
	\sum_{m^{\prime \prime}=0}^{q-1}
	c_{m^{\prime \prime}}^\dagger (\mathbf{k}_1)
	c_{m^{\prime \prime}}^\dagger (\mathbf{k}_2)
	c_{m^{\prime \prime}} (\mathbf{k}_3)
	c_{m^{\prime \prime}} (\mathbf{k}_4). \label{hint1}
\end{align}
Now, we restrict the momenta in the sum to take the values only at the dispersion minima $(0,\pi \nu/2)$, where $\nu = 0,1,2,$ or $3$.
Other momenta are also responsible for forming the momentum-space localized Wannier states, but the momentum dependence of the interaction in the momentum-space tight-binding model is correctly captured by this approximation.
We then transform $c_{m^{\prime \prime}}(\mathbf{k})$ into the band basis.
Explicitly, the transformation for $\mathbf{k} = (0,0)$ is
\begin{align}
	\begin{pmatrix}
	c_0(0,0) \\ c_1(0,0) \\ c_2(0,0) \\ c_3(0,0)
	\end{pmatrix}
	=
	\begin{pmatrix}
	\frac{\sqrt{2}+1}{2\sqrt{2}}\alpha_0 + \text{higher bands} \\
	\frac{1}{2\sqrt{2}}\alpha_0 + \text{higher bands} \\
	\frac{\sqrt{2}-1}{2\sqrt{2}}\alpha_0 + \text{higher bands} \\
	\frac{1}{2\sqrt{2}}\alpha_0 + \text{higher bands}
	\end{pmatrix}
	\approx
	\frac{1}{2\sqrt{2}}
	\begin{pmatrix}
	\sqrt{2}+1\\
	1 \\
	\sqrt{2}-1 \\
	1
	\end{pmatrix}
	\alpha_0, \label{transformation1}
\end{align}
where we ignored the contribution from the higher bands. Similarly, for the other three momenta, we obtain
\begin{align}
	\begin{pmatrix}
	c_0(0,\pi/2) \\ c_1(0,\pi/2) \\ c_2(0,\pi/2) \\ c_3(0,\pi/2)
	\end{pmatrix}
	&\approx
	\frac{1}{2\sqrt{2}}
	\begin{pmatrix}
	1 \\
	\sqrt{2}+1\\
	1 \\
	\sqrt{2}-1
	\end{pmatrix}
	\alpha_1,
	&
	\begin{pmatrix}
	c_0(0,\pi) \\ c_1(0,\pi) \\ c_2(0,\pi) \\ c_3(0,\pi)
	\end{pmatrix}
	&\approx
	\frac{1}{2\sqrt{2}}
	\begin{pmatrix}
	\sqrt{2}-1\\
	1 \\
	\sqrt{2}+1\\
	1
	\end{pmatrix}
	\alpha_2,
	&
	\begin{pmatrix}
	c_0(0,3 \pi/2) \\ c_1(0,3 \pi/2) \\ c_2(0,3 \pi/2) \\ c_3(0,3 \pi/2)
	\end{pmatrix}
	&\approx
	\frac{1}{2\sqrt{2}}
	\begin{pmatrix}
	1 \\
	\sqrt{2}-1\\
	1 \\
	\sqrt{2}+1
	\end{pmatrix}
	\alpha_3. \label{transformation2}
\end{align}
Substituting (\ref{transformation1}) and (\ref{transformation2}) into (\ref{hint1}), one obtains the momentum-space interaction Hamiltonian~(\ref{hamint}).

\end{widetext}

\end{document}